\documentclass[reprint, amsmath,amssymb, aps, prl]{revtex4-1}

\usepackage{graphicx}
\usepackage{dcolumn}
\usepackage{bm}
\usepackage{color}
\usepackage{hyperref}

\begin{document}

\preprint{APS/123-QED}

\title{Near-perfect broadband quantum memory enabled by intelligent spinwave compaction}

\author{Jinxian Guo$^{1,3}$}
\email{jxguo@sjtu.edu.cn}
\author{Zeliang Wu$^{2}$, Guzhi Bao$^{1}$, Peiyu Yang$^{1}$, Yuan Wu$^{2}$, L. Q. Chen$^{2,3}$}
\email{lqchen@phy.ecnu.edu.cn}
\author{Weiping Zhang$^{1,3,4}$}
\email{wpz@sjtu.edu.cn}

\affiliation{$^{1}$School of Physics and Astronomy, and Tsung-Dao Lee Institute, Shanghai Jiao Tong University, Shanghai 200240, P. R. China}
\affiliation{$^{2}$State Key Laboratory of Precision Spectroscopy, School of Physics and Electronics, East China Normal University, Shanghai 200062, P. R. China}
\affiliation{$^{3}$Shanghai Research Center for Quantum Sciences, Shanghai 201315, P. R. China}
\affiliation{$^{4}$Collaborative Innovation Center of Extreme Optics, Shanxi University, Taiyuan, Shanxi 030006, P. R. China}
\affiliation{$^{5}$Shanghai Branch, Hefei National Laboratory, Shanghai 201315, P. R. China}

\begin{abstract}
Quantum memory, a pivotal hub in quantum information processing, is expected to achieve high-performance storage and coherent manipulation of quantum states, with memory efficiency exceeding 90\% and quantum fidelity surpassing the non-cloning limit.
However, the current performance falls short of these requirements due to the inherent trade-off between memory efficiency enhancement and noise amplification, which not only imposes significant demands on quantum purification but also fundamentally impedes continuous-variable quantum information processing. 
In this paper, we break through these constraints, enabling high-performance quantum memory and unlocking new possibilities for quantum technologies. 
We unveil a Hankel-transform spatiotemporal mapping for light-spinwave conversion in quantum memory, and propose an intelligent light-manipulated strategy for adaptive spinwave compaction, which can maximize the conversion efficiency and simultaneously suppress the excess noise.
This strategy is experimentally demonstrated for a Raman quantum memory in warm $^{87}$Rb atomic vapor with an efficiency up to 94.6\% and a low noise level of only 0.026 photons/pulse.
The unconditional fidelity reaches 98.91\% with an average of 1.0 photons/pulse for a 17-ns input signal. 
Our results successfully demonstrate a practical benchmark for broadband quantum memory, which may facilitate advancements in high-speed quantum networks, quantum state manipulation, and scalable quantum computation.
\end{abstract}

\maketitle

\textit{Introduction.} -- The rapid advancement of quantum technologies is driving innovation in quantum computing\cite{qc1, qc2}, quantum communication\cite{qn1, qn2}, and quantum metrology\cite{app1, app2, app3, app4}. 
As these technologies shift their focus from demonstrating principles to practical applications, the performance enhancement of quantum systems has become a priority. 
Quantum memory, a crucial component for efficient and reliable quantum information processing, requires efficiency exceeding 90\% and fidelity surpassing the no-cloning limit for applications.
Currently, although quantum memory has been demonstrated to handle signal pulses from micro- to pico-seconds, covering bandwidths from kHz to THz \cite{GEM1, GEM2, EIT1, EIT2, EIT3, RM1, RM2, RM3, AFC1, AFC2}, efficiency of quantum memory with bandwidth above 10MHz still falls short of the requirement\cite{Eff2, expphase}.
Furthermore, memory efficiency, exceeding 90\% at a bandwidth of 2MHz, diminishes progressively to less than 1\% as bandwidth increases to 10GHz\cite{EIT1, RM3, AFC10}.
The low efficiency not only leads to excessive quantum error correction costs but also fundamentally obstructs continuous-variable quantum information processing.

Current approaches to enhance memory efficiency, such as increasing the power of control light and the number of atoms, often lead to a dramatic increase in excess noise, particularly four-wave mixing (FWM) noise\cite{Thomas}, which significantly reduces fidelity. 
While employing noise suppression methods\cite{phase_mismatching, noise_cavity1, noise_cavity2, Thomas, noise_polarization1, noise_polarization2, noncolinear_geometry, prllqc} often sacrifices the practical efficiency of the memory system, leading to lower efficiency.
To break this impasse, it is crucial to thoroughly comprehend the physical mechanisms underlying the trade-off between noise and efficiency in optical memory systems and explore an adequate method to achieve perfect broadband memory.

\begin{figure*}[ht]
\centering \includegraphics[scale=1]{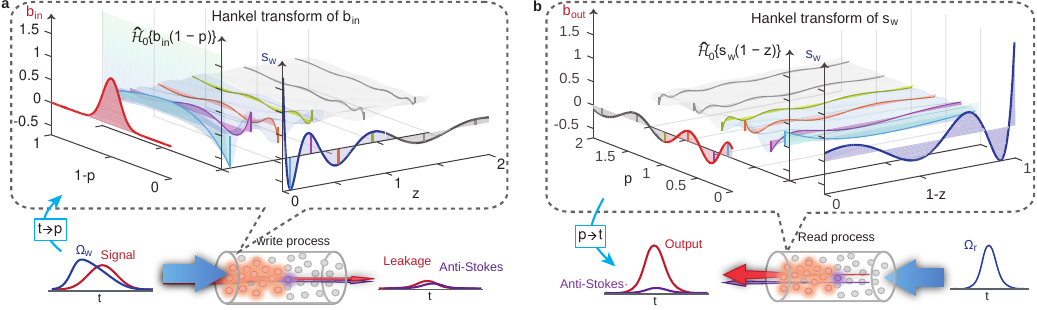}
\caption{Hankel transform mapping in Raman memory. (a) Transforming from the time t to the scaled time $p$, write process can be treated as projecting the Hankel spectral components of signal $\hat{\mathcal{H}}_{0}\{\hat{b}_{in}(1-p)\}$ onto spatial components of spinwave $s_{w}(z)$. The gray part of spinwave cannot be generated if the length of atomic ensemble limited. (b) In read process, the Hankel spectral components of spinwave $\hat{\mathcal{H}}_{0}\{\hat{s}_{w}(1-z)\}$ is projected onto temporal components of output signal $b_{out}(p)$. The gray part of output signal cannot be transformed due to the finite control power.}
\label{phy}
\end{figure*}

In this article, we unveil an intrinsic spatio-temporal Hankel-transform mapping between the temporal waveforms of lights and the spatial distribution of the spinwave in a free-space Raman protocol. 
This mapping reveals that a compact spinwave distribution is the optimal condition for achieving near-unity efficiency and minimal noise in current memory systems. 
Leveraging this mechanism, we propose an intelligent light-controlled spinwave compaction strategy, enabling optimized Hankel-transform mapping between spinwaves and light fields under practical spatio-temporal constraints. 
The intelligent modulation of light's temporal waveform generates spatially localized atomic spinwave, which enhances efficiency and reduces FWM amplification, thus effectively suppressing noise and improving fidelity.
With the strategy, we experimentally achieve a quantum memory of 17 ns (corresponding to 60 MHz bandwidth) input signal with up to 94.6$\%$ memory efficiency, 98.91$\%$ fidelity, and 0.026 photon/pulse FWM noise. 
The signal to noise ratio (SNR) in the single-photon-level experiment reaches 38.8, ensuring a noise-to-efficiency ratio of 0.028, nearly an order of magnitude improvement over the cavity enhanced case\cite{noise_cavity2}. 
This strategy, breaking the traditional trade-off between efficiency and noise, enables practical quantum memory with significant advantages of ultra-low noise, high efficiency, high fidelity, broadband, and good operability, offering new possibilities for photon state manipulation.

\textit{Protocol.} -- Our strategy can be understood qualitatively by a $\Lambda$-type free-space Raman model \cite{free, Thomas}. 
An optical signal $\hat{E}_{s,in}(t)$ is required to be coherently and efficiently mapped onto the atomic spinwave $\hat{s}$ and vice versa, driven by a time-varying strong control field with Rabi frequency $\Omega(t)$. 
However, during the mapping process, the signal light simultaneously acts as a seed, triggering a four-wave mixing amplification in the atoms by the strong control field, generating unwanted Stokes and anti-Stokes that exhibit correlated growth. 
Since the Stokes field shares the same frequency as the signal, it cannot be filtered by linear optical devices, making FWM process the dominant noise source in the high-efficiency memory process. 
Therefore, the final light field is the mixing of signal and noise, degrading the fidelity. 
This is the primary issue currently hindering efficient quantum memory.

These process can be described as
\begin{eqnarray}
\frac{\partial}{\partial z}\hat{b}&=&-g\hat{s}, \label{eq1}\\
\frac{\partial}{\partial p}\hat{s}&=&g\hat{b}. \label{eq2}
\end{eqnarray}
Here the "scaled time" \cite{lcavity} $p=\int_{-\infty}^{t}|\Omega(t')|^{2}dt'/\int_{-\infty}^{\infty}|\Omega(t')|^{2}dt'$, reflecting the dynamic atomic dispersion associated with control field. $\hat{b}=(g_{s}\hat{E}_{s}-g_{a}\hat{E}_{a}^{\dagger}e^{i\Delta kz})/(g\Omega)$ is excitation of lights consisting of the signal $\hat{E}_{s}$, the anti-Stokes noise $\hat{E}_{a}$ and the mismatched wave vector $\Delta k$ of them. 
$g=\sqrt{g_{s}^{2}-g_{a}^{2}}$ is the effective coupling coefficient where $g_{s}$ ($g_{a}$) is the coupling coefficient between spinwave $\hat{s}$ and signal(anti-Stokes).

\textit{Hankel-transform mapping.} -- By solving the equations, the mapping relation between the light excitation $\hat{b}$ and the spinwave $\hat{s}$ can be expressed by modified Hankel transform\cite{uncertainty} as 
\begin{eqnarray} 
\left(\begin{array}{c}
\hat{b}\\
\hat{s}
\end{array}\right)	=	\left(\begin{array}{cc}
1-\mathcal{\hat{H}}_{1}(p) & \mathcal{\hat{H}}_{0}(z)\\
\mathcal{\hat{H}}_{0}(p) & 1-\mathcal{\hat{H}}_{1}(z)
\end{array}\right)\left(\begin{array}{c}
\hat{b}_{in}\\
\hat{s}_{in}
\end{array}\right).
\label{sol}
\end{eqnarray}
Here $\hat{\mathcal{H}}_{m}\{f(x)\}(y)=g\int_{0}^{x}j_{m}[(x-x'),y]f(x')dx'$ is the modified Hankel transform operators with $m$-th order modified Bessel function $j_{m}(x,y)$ (see supplementary) and $x,y=p,z$. 
In the transform operators, larger $g_{s}$ and smaller $g_{a}$ can bring larger $g$, leading to the improvement of memory efficiency and the reduction of FWM noise. 
However, both $g_s$ and $g_a$ depend on the atomic density and control laser power, usually exhibiting a synchronous increase and leading to co-growth of efficiency and noise.
That is why the conventional method failed to reduce the FWM by degrading the coupling coefficient $g_a$. 
Therefore, it is necessary to find an alternative way to break the co-growth relationship between FWM noise and efficiency.

The optimization of spatial-temporal Hankel transform between light and spinwave leads to a breakthrough. 
From Eq. (3), the conversion between light and spinwave is a spatio-temporal Hankel transform $\hat{\mathcal{H}}_{0}$ while the unmapped portions, governed by $1-\hat{\mathcal{H}}_{1}$, result in leakage $\hat{b}_{l}$ in the write process and the residual spin wave $\hat{s}_{l}$ in the read process.
Ideally, the Hankel transform $\hat{\mathcal{H}}_{0}$ would be complete and the unmapped parts $1-\hat{\mathcal{H}}_{1}\sim 0$ when the atomic ensemble is infinite in length, $z\rightarrow \infty$. 
However, a practical atomic ensemble with finite length leads to imperfections in quantum memory. 
Furthermore, the FWM noise is proportional to the distribution of atomic spin wave $\hat{s}$. 
A longer spinwave distribution represents a longer propagation length of the lights in the atomic ensemble, leading to larger FWM amplification.
Therefore, optimizing the spatial-temporal Hankel transformation for a compact spinwave within the current finite-length atomic system is the key to ensure near-unity efficiency and minimal noise levels.
Below, a detailed comprehensive analysis of the Hankel-transform mapping mechanisms in the write and read processes is provided.

In the write process, the temporal spectrum of light $\hat{b}$ is mapped as the spatial distribution of spinwave $\hat{s}_{w}(z)\propto \mathcal{\hat{H}}_{0}\{\hat{b}_{in}(1-p)\}$ with a one-to-one correspondence, that is, each spectral component of $\hat{b}$ is mapped to a spinwave at a specific spatial location, as shown in Fig. \ref{phy} (a). 
Due to the finite length of the atomic ensemble, non-localized lights with high-frequency components cannot be written as a spinwave and leaked out as leakage $\hat{b}_{l}$. 
During the write process, FWM noise increases with the length of the spinwave as $\hat{E}_{a,out}^\dagger\propto-g_{a}\int_{0}^{L}e^{-i\Delta kz}\hat{s}dz$ (see supplementary for detail) where $L$ is the length of the ensemble. 
Minimizing the distribution of the generated spinwave during the writing process is crucial for suppressing FWM noise. 
Our strategy optimizes $\Omega(t)$ to ensure that the Hankel spectrum of input $\hat{b}_{in}$ is converted into a highly spatially compact spinwave $\hat{s}(z)$ within the limited ensemble length, which minimizes high-frequency components, thus achieving high efficiency, minimal leakage $\hat{b}_{l}$ and low noise simultaneously.

The read process is the reverse of the writing process, using control light $\Omega_{r}(t)$ to remap the spatial distribution of spinwave $\hat{s}_{w}$ to the output light field $\hat{b}_{out}(p)\propto \mathcal{H}_{0}\{\hat{s}_{w}(1-z)\}$ where $\mathcal{H}_{0}$ is the zeroth-order Hankel transform from space $(1-z)$ to "scaled time” $(p)$ as shown in Fig. \ref{phy} (b).
Due to the limited power of the driving field $\Omega_{r}$, the high-frequency part of spinwave's spatial Hankel spectrum cannot be converted into signal $\hat{b}_{out}$ and remain in the atomic ensemble as $\hat{s}_{l}$, limiting the retrieval efficiency to reach 100\%.
Despite increasing control power, an optimal way is to compact the spatial Hankel spectrum of $\hat{s}(1-z)$ for $\hat{b}_{out}$, ensuring a near-unity read efficiency within limited control power. 
The FWM noise in the read process is also proportional to spinwave distribution, as in the write process. This means that a compact spatial spinwave with backward retrieving can also significantly reduce FWM noise in the read process.

Therefore, the optimization of the write process is centered on manipulating the input $\hat{b}_{in}$ to create a compact spinwave $\hat{s}(z)$ within a finite atomic ensemble. 
This compact distribution also serves as the optimal spatial configuration of spinwave for the read process. 
Such an optimally compact $\hat{s}(z)$ ensures near-perfect writing and reading efficiencies while reducing noise to its minimum. 
As a result, the essence of the optimization strategy for perfect memories lies in optimizing $\hat{b}_{in}$ and its Hankel spectrum to generate a spatially compact atomic spin distribution within a limited atomic length. 
Below, we experimentally demonstrate an intelligent optimization strategy for the compact spinwave.

\begin{figure}[ht]
\centering 
\includegraphics[scale=1]{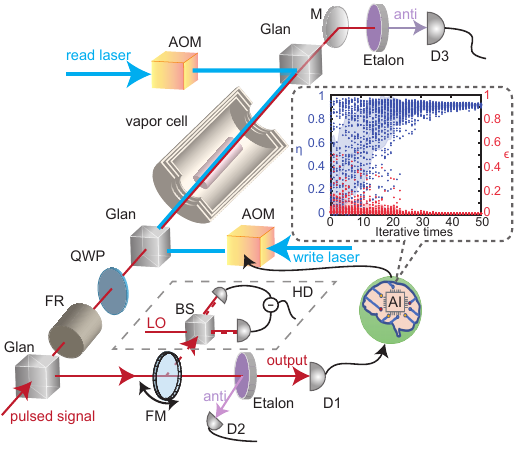}
\caption{Experiment setup of intelligent spinwave compaction strategy. The $^{87}$Rb vapor cell is heated up to 82$^{o}C$. The signal is blue-detuned by 2 GHz from the transition $|g\rangle=|{5}S_{1/2},F=2\rangle \rightarrow|e\rangle=|{5}P_{1/2},F=2\rangle$ and the write light couples to the transition $|m\rangle=|{5}S_{1/2},F=1\rangle \rightarrow|e\rangle$. The read laser is red detuned by 1.6 GHz from the transition $|m\rangle \rightarrow|e\rangle $. AOM: acousto-optic modulator, Glan: Glan Prism, QWP: quarter wave plate, FR: Faraday rotator, FM: flip mirror, D1, D2, D3: detectors. A homodyne detection (HD) has been used to estimate the quantum fidelity of memory with LO: local oscillator, BS: beam splitter. The signal from detector D1 is sent to the intelligent algorithm to generate the waveform for the AOM of the write laser. The inset shows the evolution traces when the algorithm iterates. When iterating, the memory efficiency $\eta$ (blue dots) increases while the noise ratio $\epsilon$ (red dots) decreases.}
\label{setup}
\end{figure}

\begin{figure}[ht]
\centering 
\includegraphics[scale=1]{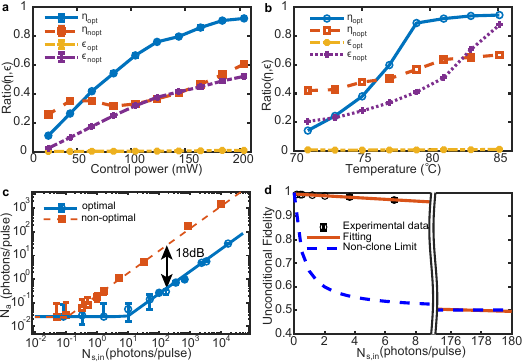}
\caption{The efficiency, noise and fidelity in experiment. Optimal efficiency $\eta_{opt}$ (blue) and non-optimal efficiency $\eta_{nopt}$ (red), alongside the optimized noise ratio $\epsilon_{opt}$ (yellow) and non-optimized noise ratio $\epsilon_{nopt}$ (purple), are shown as a function of (a) control power and (b) temperature of atomic cell. 
Here the efficiency has been modified without the amplification as $\eta=(N_{r}-N_{a})/N_{in}$ with $N_{r}$, $N_{a}$, $N_{in}$ are photon numbers of retrieved signal, anti-Stokes and input signal, respectively. 
(c) The variation of anti-Stokes photon number $N_{a}$ with the average input signal photon number $N_{s,in}$ in the optimized and non-optimized case. 
(d) The unconditional fidelity varies with averaged photon number in each pulse. As the theoretical fitting (red line) shows, the fidelity can still exceed the non-cloning limit at nearly 177 photons/pulse.}
\label{exp}
\end{figure}

\textit{Experimental setup.}--The experimental implementation of the strategy is shown in Fig. \ref{setup}. The setup includes a signal laser, which is a 17 ns-long Gaussian-shaped pulse and is two-photon resonance with a write laser.
The write light's pulse shape is modulated by an acousto-optic modulator (AOM), and after the write process, a backward-propagated retrieval light is injected into the atomic vapor for reading. Here, we adopt the backward-retrieval scheme because the commonly-used forward-readout scheme fails to find the an optimal $\Omega_{w}(t)$ that makes the spinwave $\hat{s}(z)$ in compact mode for both the writing and reading processes.
The retrieved signal is then filtered and separated using a Glan prism and a Faraday rotator. 
Efficiency $\eta=N_{out}/N_{in}$ ($N_{out}$ and $N_{in}$ are the photon numbers of the output and input signals) is measured using intensity detection and the quantum fidelity is analyzed by optical tomography. 
The anti-Stokes field generated during both the write and read process can be separated by etalons. 
The FWM process can be estimated from the emitted anti-Stokes, yielding a noise ratio $\epsilon$. 

In the experiment, the optimization of $\hat{s}(z)$ is achieved by an optimal input $\hat{b}(p)$ through precise tuning the temporal waveform of the control field $\Omega_{w}(t)$.
However, this tuning is complicated due to the nonlinear relationship from "scaled time" $p$ to real time $t$. 
To address this challenge, an intelligent spinwave compaction technology is developed, where a differential evolution (DE) algorithm\cite{alg1,alg2,alg3}, one of the intelligent algorithms, is utilized. 
According to theoretical analysis, a compact spinwave can improve efficiency while reducing noise, so the DE algorithm uses memory efficiency $\eta$ as its criterion, iterating through different temporal waveforms of $\Omega_{w}(t)$ to find the optimal one that maximizes efficiency (see supplementary).
DE algorithm continually refines and combines successful attempts, dynamically adjusting the waveform of $\Omega_{w}(t)$, and converging towards an optimal configuration. 
Throughout the iteration, a series of efficiencies $\eta$ with corresponding noise ratios $\epsilon$ have been recorded, as shown in Fig. \ref{setup} (b).
The recorded data show that the noise ratio $\epsilon$ decreases as the efficiency $\eta$ increases, aligning well with our theoretical prediction.

\textit{Experimental results.}--With the compact spinwave, the Raman quantum memory achieves high efficiency and ultra-low noise by enhancing the coupling coefficient through increased control power and atomic density. As shown in Fig. \ref{exp} (a), memory efficiency (blue) escalates with the control field power, reaching 92\%, while the noise ratio (yellow) remains minimal, consistently below 1\%, even at a control field power of 200 mW. In contrast, under the non-optimized forward-retrieval scheme, efficiency (red) remains nearly constant at 60\%, while the noise ratio (purple) surges with write power, exceeding 52\%-17.1 dB higher than our DE-optimized case.  
In Fig. \ref{exp} (b), as the temperature of the atomic vapor increases, the optimized strategy shows a slight increase in noise but a significant improvement in efficiency, reaching up to 94.6\% with a noise ratio of just 1.1\%. In contrast, the non-optimal strategy exhibits rapidly growing noise with temperature. 
Increasing the temperature of atomic cell from 81°C to 85°C improves efficiency by only 3\% but adds 37\% more FWM noise.

To further estimate the noise level of our strategy, the photon number of anti-Stokes is counted with a low-photon-level input-signal experiment.
The optimized Hankel transform (shown by the blue line in Fig. \ref{exp} (c)) guarantees a low photon number of generated anti-Stokes $N_{a}=0.026$ photon/pulse, mainly due to the spontaneous FWM process. 
In contrast, with non-optimized strategy (shown by the red line in Fig. \ref{exp} (c)), $N_{a}$ begins to increase at an input photon number of $N_{in}=0.1$ photon/pulse due to a strongly stimulated FWM process. 
The threshold of the FWM process for the optimal case is $N_{in}=10$ photon/pulse, which represents an improvement of two orders of magnitude for noise suppression.
This low noise level results in an SNR of 38.8 at the single-photon-level input signal in our protocol. 
Combining the 92\% efficiency, the noise-to-efficiency ratio (a fair metric for comparing different memory \cite{noise_cavity2,etn}) represented by $\mu_{1}=0.028$, is nearly an order of magnitude better than the reported one \cite{noise_cavity2}. 
Even with a high-photon-number input signal, our protocol achieves an 18dB noise suppression compared to the non-optimal case, ensuring an SNR of 500. 

\textit{Unconditional Fidelity.}--Unconditional fidelity is one of the core criterion for quantum memory, which should exceed the non-cloning limit of 2/3 at the single-photon level coherent field \cite{HE}. High fidelity demands both high efficiency and minimal noise, and the current scheme fulfills these essential conditions.
Using homodyne detection and quantum state tomography \cite{RM3}, the unconditional fidelity of our memory system is obtained, as shown in Fig. \ref{exp} (d). 
The unconditional fidelity reaches 98.91\% with an average of 1.0 photon per pulse in the input signal and remains at 96.82\% with an average of 6.6 photons per pulse, both of which are far beyond the non-cloning limit.
Furthermore, even when the average photon number increases to 177 photons per pulse, the fidelity still surpasses the non-cloning limit. 
This exceptional unconditional fidelity of the Raman memory indicates its significant potential for practical quantum information processing.

\textit{Conclusion.}--Our adaptive light-manipulated spinwave compaction strategy has achieved a memory efficiency of up to 94.5\% and an SNR of 38.8 corresponding a noise-to-efficiency ratio of 0.028, resulting in a single-photon-level fidelity reaching 98.91\%. 
Extending the strategy to other systems is promising, involving dynamic control of nonlinear atom-light interactions, the spatio-temporal manipulation of quantum states.
The nearly perfect quantum memory not only provides a critical element for quantum information processing but also broadens the potential of quantum communication.
Benefiting from high memory efficiency, our protocol can substantially extend the transmission distance of quantum information, especially in quantum information networks spanning over 500 km \cite{RMP}.
The high performance quantum memory may also reduce the entanglement distribution time and markedly enhances the transmission rate.
For example, in over 500km quantum key distribution that incorporate with our broadband quantum memory, the key rate may exceed tens of kilobits per seconds\cite{QKD}.
Furthermore, the high performance of our memory enables the storage of continuous-variable states, which are more sensitive to losses, thereby paving the way for continuous-variable quantum communication and quantum computation.

This work is supported by the Innovation Program for Quantum Science and Technology 2021ZD0303201, the National Science Foundation of China (Grant NO. 12234014, 11904227, 12274132,  U23A2075, 12204304, 11654005), Shanghai Municipal Science and Technology Major Project (Grant NO. 2019SHZDZX01), the Sailing Program of Shanghai Science and Technology Committee under Grant 19YF1421800, 19YF1414300, Innovation Program of Shanghai Municipal Education Commission (No.202101070008E00099), the Fundamental Research Funds for the Central Universities, the Fellowship of China Postdoctoral Science Foundation (Grant No. 2020TQ0193, 2021M702146, 2021M702150, 2021M702147, 2022T150413), and the National Key Research and Development Program of China under Grant number 2016YFA0302001. W.Z. also acknowledges additional support from the Shanghai talent program.
\nocite{*}

\bibliography{bibfile}

\end{document}